\documentclass[final]{aipproc}

\layoutstyle{6x9}

\begin{document}

\title{Absolute Polarization Measurements at RHIC in the Coulomb Nuclear
Interference Region}


\author{K.O.~Eyser}{
  address={University of California, Riverside, CA 92521, USA}}
\author{I.~Alekseev}{
  address={Institute for Theoretical and Experimental Physics (ITEP), 117259 Moscow, Russia}}
\author{A.~Bravar}{
  address={University of Geneva, 1205 Geneva, Switzerland}}
\author{G.~Bunce}{
  address={Brookhaven National Laboratory, Upton, NY 11973, USA}
  ,altaddress={RIKEN BNL Research Center, Upton, NY 11973, USA}}
\author{S.~Dhawan}{
  address={Yale University, New Haven, CT 06520, USA}}
\author{R.~Gill}{
  address={Brookhaven National Laboratory, Upton, NY 11973, USA}}
\author{W.~Haeberli}{
  address={University of Wisconsin, Madison, WI 53706, USA}}
\author{H.~Huang}{
  address={Brookhaven National Laboratory, Upton, NY 11973, USA}}
\author{O.~Jinnouchi}{
  address={KEK, Tsukuba, Ibaraki 305-0801, Japan}}
\author{Y.~Makdisi}{
  address={Brookhaven National Laboratory, Upton, NY 11973, USA}}
\author{I.~Nakagawa}{
  address={RIKEN, Wako, Saitama 351-0198, Japan}
  ,altaddress={RIKEN BNL Research Center, Upton, NY 11973, USA}}
\author{A.~Nass}{
  address={University of Erlangen, 91058 Erlangen, Germany}}
\author{H.~Okada}{
  address={RIKEN, Wako, Saitama 351-0198, Japan}
  ,altaddress={Kyoto University, Kyoto 606-8502, Japan}}
\author{E.~Stephenson}{
  address={Indiana University, Bloomington, IN 47408, USA}}
\author{D.~Svirida}{
  address={Institute for Theoretical and Experimental Physics (ITEP), 117259 Moscow, Russia}}
\author{T.~Wise}{
  address={University of Wisconsin, Madison, WI 53706, USA}}
\author{J.~Wood}{
  address={Brookhaven National Laboratory, Upton, NY 11973, USA}}
\author{A.~Zelenski}{
  address={Brookhaven National Laboratory, Upton, NY 11973, USA}}

\begin{abstract}
The Relativistic Heavy Ion Collider at Brookhaven National Laboratory provides
polarized proton beams for the investigation of the nucleon spin structure.
For polarimetry, carbon-proton and proton-proton scattering is used in the
Coulomb nuclear interference region at small momentum transfer ($-t$).
Fast polarization measurements of each beam are carried out with carbon fiber
targets at several times during an accelerator store.
A polarized hydrogen gas jet target is needed for absolute normalization over
multiple stores, while the target polarization is constantly monitored in a
Breit-Rabi polarimeter.
In 2005, the jet polarimeter has been used with both RHIC beams.
We present results from the jet polarimeter including a detailed analysis of
background contributions to asymmetries and to the beam polarization.
\end{abstract}

\maketitle


\section{Introduction}
Accurate knowledge of the beam polarization is essential in the spin program
at the Relativistic Heavy Ion Collider (RHIC).
Measurements of beam polarizations utilize scattering processes.
Fast measurements are based on proton-carbon scattering off carbon fiber
targets \cite{Nakagawa2006}, for which the necessary normalization is provided
from elastic proton-proton scattering with a polarized hydrogen gas jet
target.
At high energies, the asymmetries are driven by the electromagnetic spin-flip
amplitude at small four-momentum transfer ($10^{-3} < |-t| < 10^{-2}$).
Asymmetries are small (few percent) and contain unknown contributions from
hadronic amplitudes in this so-called Coulomb nuclear interference region.
The existing data, therefore, not only serve as a necessary input to
other RHIC experiments, but on the other hand can also be used to further
confine the helicity amplitudes $\phi_{1} - \phi_{5}$ and the hadronic
spin-flip contribution in particular \cite{Okada2006}.

%

\section{Setup and Analysis}
\begin{figure}
  \includegraphics[height=.42\textwidth]{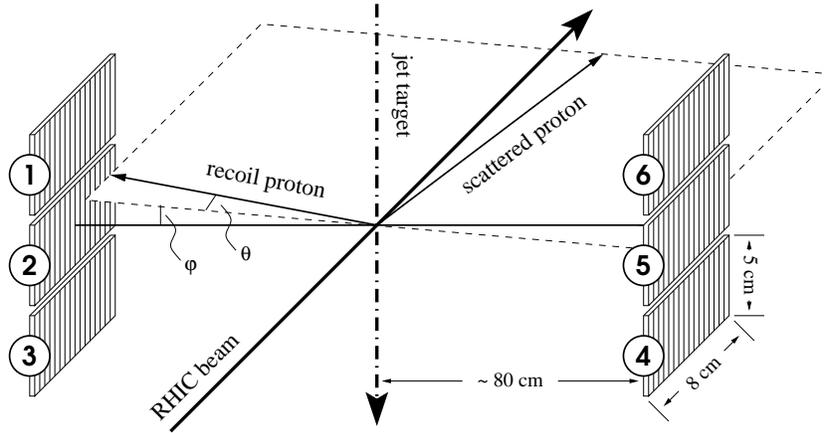}
  \caption{\label{jetpol}
  Detector setup of the jet polarimeter at RHIC. The six detectors are
  centered with respect to the interaction point. In routine operation only
  one RHIC beam hits the target, while the other beam is displaced.}
\end{figure}

The jet polarimeter is located at one of the collision points in the RHIC
accelerator and reflects the kinematics of elastic proton-proton scattering at
small $|-t|$.
It consists of a polarized hydrogen atomic jet target with a Breit-Rabi
polarimeter \cite{Gaul1992} and a set of six separate detectors each of 16
silicon strips, see figure \ref{jetpol}.
The target polarization is prepared in an inhomogeneous magnetic field in
combination with a radio-frequency transition unit and constantly monitored
(not shown in the figure).
The hydrogen molecular content has been determined and is corrected for.
The effective polarization amounts to $P_{target}=(92.4 \pm 1.8)\%$.

The detectors are centered with respect to the interaction point of the proton
beam with the target.
In this geometry, eight downstream strips on each detector can
detect elastically scattered recoil protons when one RHIC beam hits the
target.
In routine operation, one of the two RHIC beams is displaced, so the
respective eight non-signal strips can be used to estimate the background
fraction below the elastic signal peak.

The silicon detectors are read out with waveform digitizers (running at 420
MHz and synchronized with the RHIC clock) that send the pre-processed ADC
spectra to the DAQ-PC.
Two different $\alpha$-sources (Am and Gd) are used for energy calibration and
estimation of the entrance window thickness of the detectors.
Additional time of flight offsets are individually adjusted for each strip
using the pronounced proton signal.
Particle identification is based on time of flight and energy.
Elastic scattering further correlates the detector geometry, i.e.\ the
scattering angle, and the small recoil energy of the proton.
A time-of-flight cut of a few ns removes the major part of prompt
background events below 5 MeV.

In 2005, one of the two RHIC beams was centered on the jet target for several
days to accumulate enough statistics for a precise measurement of the beam
polarization.
Both beams have been measured repeatedly over the course of a few weeks.

For the determination of the RHIC beam polarization, a set of four
different vertical polarization combinations of target and beam is used.
The yields are then combined to separate asymmetries resulting from the beam
and target polarizations.
While a certain polarization direction would result in a specific left-right
asymmetry in the detectors, respective yields can be coupled with their
opposite polarization directions in order to remove detector acceptances and
efficiencies and luminosity effects, \cite{Ohlsen1973}.
The beam polarization $P_{beam}$ is then derived from the ratio of the
measured asymmetries $\epsilon_{beam}$ and $\epsilon_{target}$ and the known
target polarization $P_{target}$:
\begin{equation}
  P_{beam} = - \frac{\epsilon_{beam}(T_{R})}{\epsilon_{target}(T_{R})}\cdot P_{target}.
\end{equation}
The analyzing power $A_{N} = \epsilon / P$ is not flat in the considered
$t$-region and, therefore, the beam polarization was determined in separate
steps of the recoil energy $T_{R}$ before calculating the weighted mean.
Also, the kinematic correlation of the detector strip position and the recoil
energy suppresses the background outside of the considered energy range, where
no elastic signal is seen.

The ratio of asymmetries is very robust with regards to background
contributions to the yields, as long as the background has no pronounced
polarization dependence.
Background fractions can be as large as 7\% in certain energy ranges, which
affect the beam polarization only in second order.

\section{Results}

\begin{figure}
  \includegraphics[height=.42\textwidth]{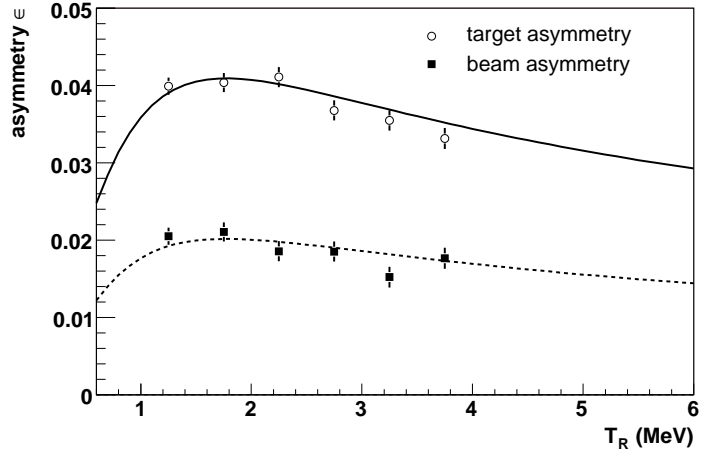}
  \caption{\label{results}
    Target and beam asymmetries as functions of recoil energy $T_{R}$.
    The solid and dashed lines are expected asymmetries from formal
    descriptions in terms of helicity amplitudes \cite{Okada2006} in the
    Coulomb nuclear interference region scaled with the polarization values.}
\end{figure}

Target and beam asymmetries have been measured as functions of energy of the
recoil proton.
Figure \ref{results} shows the typical peak of the asymmetries (and the
analyzing power) at $T_{R} \approx 1.5$ MeV in the CNI region from a subset of
the data.
The asymmetry ratio between 1.0 and 4.0 MeV is used to determine the beam
polarization.
Lower and higher recoil energies are discarded due to increased acceptance
asymmetries and highly increased background.
Also, the asymmetries in figure \ref{results} are compared to the expected
shape of the analyzing power $A_{N}$ from a formal description in terms of
helicity amplitudes \cite{Okada2006} that has been scaled with the target and
beam polarizations.
The shown curves are neither fitted to the analyzed data, nor do they include
contributions from hadronic spin-flip amplitudes.

In 2005, beam polarizations of nearly 50\% have been measured in RHIC with
good statistical accuracy.
Systematic errors were estimated from the background yields from
empty target measurements and subdivision with respect to separate proton
bunches in the accelerator.
Other checks included many thousand repetitions of random assignments to the
polarization directions of bunches and calculating the resulting
asymmetries.

Background contributions have been identified from molecular content in the
jet target, from beam gas, and from the displaced beam that is threaded around
the target.
No significant beam or target polarization dependence has been observed in the
background, which is uniformly distributed over the detector halves within the
statistical uncertainties.
The effect on the determination of the beam polarization has been estimated by
purposefully widening the signal region on the detectors, this way increasing
the background by factors of up to four and consecutively lowering the beam
and target asymmetries.
The ratio of asymmetries, on the other hand, is largely unaffected by the
growing background and an upper limit of $\Delta P / P = 1.1\%$ has been
assigned to the beam related background below the signal.
The hydrogen molecular content of the target causes an additional uncertainty
that is already taken into account in the target polarization error.


\section{Summary}
Fast beam polarization measurements at RHIC are carried out with two carbon
polarimeters \cite{Nakagawa2006}.
Absolute polarization normalization is done alternatingly with a polarized
hydrogen jet target.
Beam polarizations of nearly 50\% have been measured in the latter half of the
2005 run.
Total uncertainties have reached the accuracy that experiments in the RHIC
spin programm have called for.
The existing data, also, have been used to further constrain the knowledge of
hadronic spin-flip amplitudes at $\sqrt{s}=13.7$ and $6.9$ GeV
\cite{Okada2006}.


\bibliographystyle{aipproc}   

\end{document}